\documentclass[10pt]{article}

\usepackage{amsmath}
\usepackage{amssymb}
\usepackage{graphics}
\usepackage{rotating}
\usepackage{cite}
\usepackage{color}
\usepackage{fancybox}
\usepackage{pstricks}
\usepackage{xcolor}
\usepackage{shadowtext}
\usepackage{multicol}
\def\0{\mbox{\tiny $0$}}
\def\1{\mbox{\tiny $1$}}
\def\2{\mbox{\tiny $2$}}
\def\3{\mbox{\tiny $3$}}
\def\4{\mbox{\tiny $4$}}
\def\5{\mbox{\tiny $5$}}
\def\6{\mbox{\tiny $6$}}
\def\7{\mbox{\tiny $7$}}
\def\8{\mbox{\tiny $8$}}
\def\9{\mbox{\tiny $9$}}
\definecolor{navy}{rgb}{0,0,.6}
\definecolor{jour}{rgb}{0,0.6,.4}
\definecolor{jbul}{rgb}{0.7,0.,.4}
\begin{document}
%
%%%%%%%%%%%%%%%%%%%%%%%%%%%%%%%%%%%%%%%%%%%%%%%%%%%%%%%%%%%%%%%
\thispagestyle{empty}
\setcounter{page}{0}

\begin{center}
\shadowrgb{0.8,0.8,1}
\shadowoffset{4pt}
\shadowtext{
\color{navy}
\fontsize{18}{18}\selectfont
\bf THE FREQUENCY CROSSOVER FOR}\\
\shadowtext{
\color{navy}
\fontsize{18}{18}\selectfont
\bf THE GOOS-H\"ANCHEN SHIFT}
\end{center}

\vspace*{1cm}

\begin{center}
\shadowrgb{0.8, .8, 1}
\shadowoffset{2.5pt}
\shadowtext{\color{jbul}
\fontsize{13}{13}\selectfont
$\boldsymbol{\bullet}$}
\shadowtext{\color{jour}
\fontsize{15.5}{15.5}\selectfont
\bf Journal of Modern Optics 60, 1772-1780 (2013)}
\shadowtext{\color{jbul}
\fontsize{13}{13}\selectfont
$\boldsymbol{\bullet}$}
\end{center}

\vspace*{1cm}

\begin{center}
\begin{tabular}{cc}
\begin{minipage}[t]{0,5\textwidth}
{\bf Abstract}.
For total reflection, the Goos-H\"anchen (GH) shift is proportional to the wavelength of the laser beam. At critical angles, such a shift is instead proportional to the square root of the product of the beam waist and  wavelength. By using the stationary phase method (SPM) and, when necessary, numerical calculations, we present a detailed analysis of the frequency crossover for the GH shift. The study, done in different incidence regions, sheds new light on the validity of the analytic formulas found in literature.
\end{minipage}
& \begin{minipage}[t]{0,5\textwidth}
{\bf Manoel P. Araujo}\\
Gleb Wataghin Physics Institute\\
State University of Campinas (Brazil)\\
{\color{navy}{{\bf mparaujo@ifi.unicamp.br}}}
\hrule
\vspace*{0.2cm}
{\bf Silv\^ania A. Carvalho}\\
Department of Applied Mathematics\\
State University of Campinas (Brazil)\\
{\color{navy}{{\bf  silalves@ime.unicamp.br}}}
\hrule
\vspace*{0.2cm}
{\bf Stefano De Leo}\\
Department of Applied Mathematics\\
State University of Campinas (Brazil)\\
{\color{navy}{{\bf  deleo@ime.unicamp.br}}}
\end{minipage}
\end{tabular}
\end{center}

\vspace*{2cm}

\begin{center}
\shadowrgb{0.8, 0.8, 1}
\shadowoffset{2pt}
\shadowtext{\color{navy}
{\bf
\begin{tabular}{ll}
I. & INTRODUCTION \\
II. & THE INCOMING GAUSSIAN BEAM \\
III. & THE SPATIAL PHASE OF THE OUTGOING BEAM \\
IV.  & TRANSMISSION COEFFICIENT AND GEOMETRICAL PATH\\
V. & THE GOOS-H\"ANCHEN SHIFT BY THE SPM\\
VI. & NUMERICAL ANALYSIS\\
VII. & CONCLUSIONS\\
& \\
& \,[\,14 pages, 4 figures\,]
\end{tabular}
}}
\end{center}

\vspace*{7cm}

\begin{flushright}
\shadowrgb{.8, .8, 1}
\shadowoffset{2pt}
\shadowtext{\color{jbul}
\fontsize{11}{11}\selectfont
$\boldsymbol{\bullet}$}
\hspace*{-.2cm}
\shadowtext{
\color{jour}
\fontsize{15.5}{15.5}\selectfont
$\boldsymbol{\Sigma\hspace*{0.06cm}\delta\hspace*{0.035cm}\Lambda}$}
\hspace*{-.35cm}
\shadowtext{\color{jbul}
\fontsize{11}{11}\selectfont
$\boldsymbol{\bullet}$}
\end{flushright}

%%%%%%%%%%%%%%%%%%%%%%%%%%%%%%%%%%%%%%%%%%%%%%%%%%%%%%%%%%%%%%%%%%
%%%%%%%%%%%%%%%%%%%%%%%%%%%%%%%%%%%%%%%%%%%%%%%%%%%%%%%%%%%%%%%%%%%%%%%

\newpage

%%%%%%%%%%%%%%%%%%%%%%%%%%%%%%%%%%%%%%%%%%%%%%%%%%%%%%%%%%%%%%%%%%%%%%%%%%
%%%%%%%%%%%%%%%%%%%%%%%%%%%%%%  SECTION I    %%%%%%%%%%%%%%%%%%%%%%%%%%%%%
%%%%%%%%%%%%%%%%%%%%%%%%%%%%%%%%%%%%%%%%%%%%%%%%%%%%%%%%%%%%%%%%%%%%%%%%%%

\section*{\normalsize I. INTRODUCTION}

The GH shift, widely investigated in the last decades, continues to attract attention due
to available technologies \cite{2000JOSAA17,2002JOSAA19,1988JOSAA5,1983JAP54,2010JOSAB27,2012JOSAA29,2012OE20,2012NJP14,1977JOSA67p103,2012NJP14p013058,2013JO15p014001,2013JO15p014009}. This phenomenon refers to the lateral shift of a totally reflected beam with respect to the optical path expected from geometrical optics.  For an interesting overview of the effect and its generalizations, we suggest  the references \cite{2013JO15p014001,2013JO15p014009}. The fact that total reflection does not take place at the spatial point predicted by the geometrical optics, see Fig\,1a, was discovered by Newton \cite{1718WI1730} which proposed that the path during total reflection is a parabola, the vertex being within the rarer medium. The problem was then experimentally and theoretically analyzed  by other authors in 1920s and 1930s \cite{1925AP382,1937AP422}. The first experimental observation of this effect was done by Goos and H\"anchen \cite{1947AP436}. Their experiment stimulated the study of this shift for different polarized electromagnetic waves \cite{1948AP437,1949AP439}. In the literature, the longitudinal and transverse shifts of an optical beam have been investigated by using the far-field measurement for the reflected field
and the scanning tunneling optical microscope for the (evanescent) transmitted beam \cite{2000JOSAA17}.

The fact that the GH shift has been investigated in different physical problems, see frustrated total internal reflection (FTIR) \cite{1977JOSA67,1986OC56,2001PRE63}, partial reflection \cite{2003PRL91}, acoustic \cite{2000JASA108}, nonlinear optics \cite{1995PRL75}, frequency dispersive media\cite{REF1,REF2},  and surface physics \cite{1960PRL4}, shows the great and increasing attention of the scientific community to this topic. Some interesting  applications, which include sensors and optical waveguide switching, can be found in recent research works \cite{2012NJP14,2006APL89,2008OL33,2000APL76}. On the other hand, the measurement of this very small shift (of the order of the beam wavelength)  represented a continuous challenge. The proposed solutions to increase its magnitude were discussed in many works \cite{2012OE20,2007OL32,1993PRL70}. Over the years, this technical problem was solved and it is now possible to measure this shift\cite{2011OptExp19p9636}.

In 1948, Artmann obtained an expression for the GH shift at critical angle \cite{1948AP437}. After few decades, other authors studied the same problem in detail obtaining an expression for the lateral shift for incidence angles very close to the critical angle \cite{1986JOSAA3,1988JOSAA5p132,1971JOSA61}.

In this paper, we describe the beam propagation into the dielectric block by using the analogy between optcis and  quantum mechanics\cite{2011EPJD61p481,2011EPJD65p563}. In particular, we derive the transmission and reflection coefficients at each interface and then  calculate both analytically and numerically the position of the outgoing beam. The wave packet formalism is introduced by considering a gaussian wave number distribution. The SPM,  applied to our gaussian beams, leads to an analytical formula for the exit point of the optical beam propagating throughout a right angle prism with refractive index $n$ \cite{1955PR98,1975DOVER,2008JOA10}. The  difference between the geometrical result obtained by the Snell law and the real optical path and its frequency crossover will be the subject of our investigation.

The study done in this paper allows to investigate the validity of the analytical formulas for different incidence angles as well the transition from partial to total internal reflection. A new formula, based on the SPM, is presented for the shift at critical incidence.  The numerical analysis confirms our analytical predictions.
%Our analytical curve could not be complete because a rigorous calculation has to been done to complete this results.

This paper is organized as follows. In the following Section, we discuss the Gaussian beam propagation in  free space. The geometry of the physical system is presented in Section III. In Section IV, we calculate, by using the analogy between optics and quantum mechanics, the reflection and transmission coefficients for  $s$ and $p$ polarized waves\cite{1999WOLF}. In section V, by using the SPM, we analyze the quantum additional phase and give our analytical expressions for the GH shift for different incidence regions. The numerical results are then presented in section VI. Our conclusions and proposals  for future investigations are drawn in the final section.

\section*{\normalsize II. THE INCOMING GAUSSIAN BEAM}

The beam propagation in free space is determined by the wave number laser distribution\cite{2007SALEH}. Let us consider the following distribution,
\begin{equation}
G(\boldsymbol{k})= \exp \left[-\frac{(k_{x}^{^2} +
k_{y}^{^2})\,\mbox{w}_{\0}^{\2}}{4}\right]\,\,  \delta \left(k_{z} -
\sqrt{k^{^{2}} - k_{x}^{^2} - k_{y}^{^2}}\right),
\end{equation}
\noindent where $\mbox{w}_{\0}$  is the beam waist size  and $k = 2\pi / \lambda$. The electric field amplitude is then given by
\begin{eqnarray}
E_{_{in}}(\boldsymbol{r}) &=& E_{\0} \frac{\mbox{w}_{\0}^{\2}}{4
\pi}\,\int\hspace*{-.1cm} \mbox{d} \boldsymbol{k}\,\, G(\boldsymbol{k})\,  \,
\exp
\left[\,i\,\boldsymbol{k}\cdot\boldsymbol{r}\,\right], \nonumber \\
&=& E_{\0} \frac{\mbox{w}_{\0}^{\2}}{4 \pi}\,\int\hspace*{-.1cm}
\mbox{d}k_x\,\mbox{d}k_y\,\, \exp \left[-\frac{(k_{x}^{^2} +
k_{y}^{^2})\,\mbox{w}_{\0}^{\2}}{4}\right]\,  \, \exp
\left[\,i\,\left(k_x\,x+k_y\,y+\sqrt{k^{^{2}} - k_{x}^{^2} -
k_{y}^{^2}}\,z\right)\right],
\end{eqnarray}

\noindent with $E_{\0} = E (\boldsymbol{0})$. By considering the paraxial approximation (valid for $k\,\mbox{w}_{\0}\geq  5$), the integrations can be done analytically leading to
\begin{eqnarray}
E_{_{in}}(\boldsymbol{r}) &\approx & E_{\0} \,\frac{\mbox{w}_{\0}^{\2}}{4
\pi}\,\,e^{ik\,z}\, \int\hspace*{-.1cm} \mbox{d}k_x\,\mbox{d}k_y\,\, \exp
\left[-\frac{(k_{x}^{^2} + k_{y}^{^2})\,\mbox{w}_{\0}^{\2}}{4}\right]\,  \,
\exp \left[\,i\,\left(k_x\,x+k_y\,y - \frac{k_{x}^{^2} +
k_{y}^{^2}}{2\,k}\,z\right)\right]\nonumber \\
 & \approx & E_{\0}\, \frac{\mbox{w}_{\0}^{\2}}{\mbox{w}_{\0}^{\2}+2\,i\,
 \displaystyle{\frac{z}{k}}}\,\,e^{ik\,z}\,\,\exp \left[-\frac{x^{\2} +
y^{\2}}{\mbox{w}_{\0}^{\2}+2\,i\,\displaystyle{\frac{z}{k}} }\right].
\label{eq:Ein}
\end{eqnarray}
 The intensity, $I(\boldsymbol{r}) = \left|E(\boldsymbol{r})\right|^{^2}$, for the incoming beam is then given by
\begin{equation}
I_{_{in}}(\boldsymbol{r}) \,\approx \,\,  \mbox{I}_{\0} \,\left[\,
\frac{\mbox{w}_{\0}}{\mbox{w}(z)}\,\right]^{^{2}}\,\,\exp
\left[-\,2\,\frac{x^{\2} + y^{\2}}{\mbox{w}^{\2}(z)}\right],
\end{equation}
\noindent where $\mbox{w}(z) = \mbox{w}_{\0}\, \displaystyle{\sqrt{1+\left(\, \frac{\lambda\,z}{\pi\,\mbox{w}_{\0}^{^{2}}}\,\right)^{^{2}}}}$.

\section*{\normalsize III. THE SPATIAL PHASE OF THE OUTGOING BEAM}

In this section, we determine the direction of propagation of the outgoing beam by calculating its spatial phase. In doing this, it is convenient to introduce new axes of coordinates, see Fig.\,1b. The plane of incidence is the $y$-$z$ plane, where $z$ is the propagation direction of the incoming beam, and the new axes represent rotation of the $y$-$z$ system with $z_{_{in}}$,  $z_{_{*}}$, and $z_{_{out}}$ normal to the left (air/dielectric), down (dielectric/ar) and right (dielectric/air) interface. Denoting the incidence angle by $\theta$ and using $R(\theta)$ to identify a counterclockwise,
\begin{equation*}
R(\theta) = \left( \begin{array}{cc}
\cos\theta & -\sin\theta \\
\sin\theta & \cos\theta \end{array} \right)\,\,,
\end{equation*}
we find
\begin{equation}
\label{rot}
\left( \begin{array}{c}
y_{_{out}} \\
z_{_{out}} \end{array} \right) =  R \left(\frac{3\,\pi}{4}\right) \left(
\begin{array}{c}
y_{_*} \\
z_{_*} \end{array} \right) = R\left( \frac{\pi}{2} \right) \left(
\begin{array}{c}
y_{_{in}} \\
z_{_{in}} \end{array} \right) = R\left(\frac{\pi}{2} - \theta \right) \left(
\begin{array}{c}
y \\
z \end{array} \right).
\end{equation}
The spatial phase of the incoming beam is
\begin{equation}
\varphi_{_{in}}=\,\,\boldsymbol{k}\,\,\cdot\,\,\boldsymbol{r} \,\,=\,\, \boldsymbol{k}_{_{in}} \cdot\,\,\boldsymbol{r}_{_{{in}}}\,\,,
\end{equation}
where $\boldsymbol{r}_{_{{in}}}=(x\,,y_{_{in}},z_{_{in}})$ is obtained from Eq.\,(\ref{rot}) and
$\boldsymbol{k}_{_{in}}$ is given by
\begin{equation}
k_{x_{_{in}}}=k_x\,\,\,\,\,\,\,\mbox{and}\,\,\,\,\,\,\,\left( \begin{array}{c}
k_{y_{_{in}}} \\
k_{z_{_{in}}} \end{array} \right)  =  R(-\theta) \left( \begin{array}{c}
k_y \\
k_z \end{array} \right).
\end{equation}
Taking into account that the discontinuity is along the $z_{_{in}}$ axis, the  $x_{_{in}}(=x)$  and $y_{_{in}}$ components of the wave number do not  change when the beam crosses the first air/dielectric boundary,
 \begin{equation} (q_{x}\,,q_{y_{_{in}}})=(k_x\,,k_{y_{_{in}}})\,\,\,\,\,\,\,\,\Rightarrow\,\,\,\,\,\,\,\,
q_{z_{_{in}}} = \sqrt{n^{\2}k^{^{2}} - k_{x}^{^2} -k_{y_{_{in}}}^{^2}}\,\,.
\end{equation}
 The spatial phase of the  beam moving in the dielectric from the left (air/dielectric) to the down (dielectric/air) interface is
 \begin{equation}
 \label{spa12}
\varphi_{_{left/down}}= \,\,\boldsymbol{q}_{_{in}}\cdot\,\,\boldsymbol{r}_{_{in}} =\,\, \boldsymbol{q}_{_*} \cdot\,\,\boldsymbol{r}_{_{*}}\,\,,
 \end{equation}
 where  $\boldsymbol{r}_{_{{*}}}=(x\,,y_{_{*}},z_{_{*}})$ is obtained from Eq.\,(\ref{rot}) and
$\boldsymbol{q}_{_{*}}$ is given by
\begin{equation}
q_x{_{_{*}}}=k_x\,\,\,\,\,\,\,\mbox{and}\,\,\,\,\,\,\,\left(
\begin{array}{c}
q_{y_{_{*}}} \\
q_{z_{_{*}}} \end{array} \right)  =  R \left( -\frac{\pi}{4}\right) \left(
\begin{array}{c}
q_{y_{_{in}}} \\
q_{z_{_{in}}} \end{array} \right)\,\,.
\end{equation}
The spatial phase of the reflected beam at the down interface is obtained from Eq.\,(\ref{spa12}) by changing $z_{_{*}}$ in  $-\,z_{_{*}}$,
\begin{equation}
\label{spa23}
\varphi_{_{down/right}}=\,\,
q_{x_{_*}}x_{_{*}}+\,q_{y_{_*}}y_{_{*}}-\,q_{z_{_*}}z_{_{*}} =\,\, \boldsymbol{q}_{_{out}}\cdot\,\,\boldsymbol{r}_{_{out}}\,\,,
\end{equation}
where $\boldsymbol{r}_{_{{out}}}=(x\,,y_{_{out}},z_{_{out}})$ is obtained from Eq.\,(\ref{rot}) and
$\boldsymbol{q}_{_{out}}$ is given by
\begin{equation}
q_x{_{_{out}}}=k_x\,\,\,\,\,\,\,\mbox{and}\,\,\,\,\,\,\,\left(
\begin{array}{c}
q_{y_{_{out}}} \\
q_{z_{_{out}}} \end{array} \right)  =  R \left( \frac{3\,\pi}{4}\right) \left(
\begin{array}{r}
q_{y_{_{*}}} \\
-\,q_{z_{_{*}}} \end{array} \right) =\left(
\begin{array}{r}
-\,q_{y_{_{in}}} \\
q_{z_{_{in}}} \end{array} \right).
\end{equation}
The beam reaches the right (dielectric/air) boundary and due to the fact that the discontinuity is along the $z_{_{*}}$ axis, the  $x_{_{*}}(=x)$  and $y_{_{*}}$ components of the wave number do not  change when the beam crosses the last dielectric/air interface,
 \begin{equation} (k_{x}\,,k_{y_{_{out}}})=(k_x\,,q_{y_{_{out}}})=(k_x\,,-\,q_{y_{_{in}}})=(k_x\,,-\,k_{y_{_{in}}})\,\,\,\,\,\,\,\,\Rightarrow\,\,\,\,\,\,\,\,
k_{z_{_{out}}} =k_{z_{_{in}}} \,\,.
\end{equation}
Finally, the spatial phase of the outgoing beam is
\begin{eqnarray}
\varphi_{_{out}}=\,\,
\boldsymbol{k}_{_{out}} \cdot\,\,\boldsymbol{r}_{_{{out}}}\,\,& = &\,\,k_x\, x \,-\, k_{y_{_{in}}} y_{_{out}} +\,\, k_{z_{_{zin}}} z_{_{out}}\nonumber \\
 & = & k_x\, x + (\,k_z\cos 2\theta - k_y\sin 2 \theta\,)\,y +(\,k_z\sin 2\theta + k_y \cos 2 \theta\,)\,z\,\,.
\end{eqnarray}
Consequently, the outgoing wave number vector in the $(x,y,z)$ coordinates system is given by
\begin{equation}
\left[\,\nabla \varphi_{_{out}}\,\right]_{_{(k_x=0,k_y=0)}} = (\,0\,,\,k\cos 2\theta\,,\,k\sin
2\theta\,)\,\,\,\,\,\left\{\begin{array}{lcl}  (\,0\,,\,0\,,\,k\,) & & \mbox{for}\,\,\,\,\,\theta=\pi/4\,\,,   \\ (\,0\,,\,k\,,\,0\,) & & \mbox{for}\,\,\,\,\,\theta=0\,\,,  \\
 (\,0\,,\,0\,,\,-\,k\,) & & \mbox{for}\,\,\,\,\,\theta=-\,\pi/4\,\,.
\end{array} \right.
\end{equation}
The outgoing beam propagates parallel to the incoming beam for $\theta=\pm\, \pi/4$ and perpendicular to the incoming beam for $\theta=0$.

\section*{\normalsize IV. TRANSMISSION COEFFICIENT AND GEOMETRICAL PATH}

The Fresnel formulas for the reflection and transmission coefficients of $s$-polarized beam can be
given by using the analogy between optics and quantum mechanics\cite{2008JOA10}. From the non-relativistic quantum analysis of the step potential \cite{1999WOLF},
\begin{equation}
r[\alpha, \beta] =\frac{\alpha - \beta} {\alpha + \beta}\,\,\,\,\,\,\,\mbox{and}\,\,\,\,\,\,\,
t[\alpha, \beta]  = \frac{2\, \alpha}{\alpha + \beta}\,\,,
\end{equation}
we find
\begin{equation}
\begin{array}{lll}
r_{_{in}}^{^{(s)}} = r[k_{z_{_{in}}}, q_{z_{_{in}}}]\,\exp[\,2\,i\,k_{z_{_{in}}} a_{_{in}}] &\,\,,\,\,\,\,\,\,\, &  t_{_{in}}^{(s)} = t[k_{z_{in}}, q_{z_{in}}]\, \exp[\,i\,(k_{z_{_{in}}}-\,q_{z_{_{in}}})\, a_{_{in}}]
\\ \\
r_{_{*}}^{^{(s)}} = r[q_{z_{_{*}}}, k_{z_{_{*}}}]\,\exp[\,2\,i\,q_{z_{_{*}}} a_{_{*}}] &
 \,\,,\,\,\,\,\, & t_{_{*}}^{(s)} = t[q_{z_{*}}, k_{z_{*}}]\, \exp[\,i\,(q_{z_{_{*}}}-\,k_{z_{_{*}}})\, a_{_{*}}]\\ \\
 r_{_{out}}^{^{(s)}} = r[q_{z_{_{out}}}, k_{z_{_{out}}}]\,\exp[\,2\,i\,q_{z_{_{out}}} a_{_{out}}] &
 \,\,,\,\,\,\,\, & t_{_{*}}^{(s)} = t[q_{z_{out}}, k_{z_{out}}]\, \exp[\,i\,(q_{z_{_{out}}}-\,k_{z_{_{out}}})\, a_{_{out}}]\,\,.
\end{array}
\label{eq:coef}
\end{equation}
By choosing the axes origin in $a_{_{in}}=0$ (this implies $a_{_{*}}=a/\sqrt{2}$\,\, and $a_{_{out}}=b-a$)
and observing that $q_{z_{_{out}}}=q_{z_{_{in}}}$ and $k_{z_{_{out}}}=k_{z_{_{in}}}$, we obtain the following expression for the transmission coefficient
 \begin{equation}
t^{^{(s)}} = t_{_{in}}^{^{(s)}} \,\, r_{_{*}}^{^{(s)}} \,\, t_{_{out}}^{^{(s)}} = \frac{4 \, k_{z_{_{in}}} q_{z_{_{in}}}}{(k_{z_{_{in}}}+ q_{z_{_{in}}})^{^2}}\, \frac{q_{z_{_*}}-\,k_{z_{_*}}}{q_{z_{_*}}+\, k_{z_{_*}}} \,\,
\exp[\,i\,\psi_{_{out}}\,]\,\,,
\label{eq:eq17x}
\end{equation}
where
\begin{equation}
\psi_{_{out}} = q_{z_*} a\sqrt{2} +  (q_{z_{_{in}}}-\,k_{z_{_{in}}})\,(b-a)\,\,.
\end{equation}
The transmission coefficient for $p$-polarized is found by the substitution rule
\[ (\,k_{z_{in,*}}\,,\,q_{z_{in,*}}\,)\,\,\,\,\,\to\,\,\,\,\,\, \left(\,n\,k_{z_{in,*}}\,,\,\frac{q_{z_{in,*}}}{n}\,\right)\,\,,  \]
in
\[
\frac{4 \, k_{z_{_{in}}} q_{z_{_{in}}}}{(k_{z_{_{in}}}+ q_{z_{_{in}}})^{^2}}\, \frac{q_{z_{_*}}-\,k_{z_{_*}}}{q_{z_{_*}}+\, k_{z_{_*}}} \,\,.
\]
This leads to
\begin{equation}
t^{^{(p)}} = \frac{4 \, n^{\2}k_{z_{_{in}}} q_{z_{_{in}}}}{(n^{\2}k_{z_{_{in}}}+ q_{z_{_{in}}})^{^2}}\, \frac{q_{z_{_*}}-\,n^{\2}k_{z_{_*}}}{q_{z_{_*}}+\, n^{\2}k_{z_{_*}}} \,\,
\exp[\,i\,\psi_{_{out}}\,]\,\,.
\end{equation}
Once obtained the transmission coefficient, we can write, by using a  gaussian convolution, the amplitude of the outgoing electric field\cite{2008JOA10},
\begin{eqnarray}
E^{^{(s,p)}} _{_{out}}(\boldsymbol{r}) &=& E_{\0} \frac{\mbox{w}_{\0}^{\2}}{4 \pi}\,\int\hspace*{-.1cm}
\mbox{d}k_x\,\mbox{d}k_y\,\, \,t^{^{(s,p)}} \exp \left[-\frac{(k_{x}^{^2} +
k_{y}^{^2})\,\mbox{w}_{\0}^{\2}}{4}\right]\,  \, \exp
\left[\,i\,\boldsymbol{k}_{_{out}} \cdot\,\,\boldsymbol{r}_{_{out}}\,\right]\,\,.
\end{eqnarray}
The geometrical path can be now calculated by using the SPM \cite{1955PR98,1975DOVER,2008JOA10}. To illustrate the method, let us consider the incoming beam. By imposing that the derivative of  the phase  is zero at the center of our symmetric gaussian distribution, we immediately find
\begin{equation}
\left[\frac{\partial \varphi_{_{in}}}{\partial {k_{x}}},\frac{\partial \varphi_{_{in}}}{\partial {k_{y}}}\right]_{_{(k_x=0,k_y=0)}} = (0,0)\,\,\,\,\,\,\,\,\,\,\,\,\Rightarrow\,\,\,\,\,\,\,\,\,\,\,\,(\,x_{_{\mbox{\tiny max}}}\,,\,y_{_{\mbox{\tiny max}}}\,)_{_{in}}=
(\,0\,,\,0\,)\,\,.
\end{equation}
For the outgoing beam, the phase is given by the spatial phase $\varphi_{_{out}}$ and by the phase  $\psi_{_{out}}$ coming from the transmission coefficient. By using the SPM, we find
\begin{equation}
\left[\frac{\partial (\varphi_{_{out}}+ \psi_{_{out}})}{\partial {k_{x}}}\right]_{_{(0,0)}} = 0\,\,\,\,\,\,\,\,\,\,\,\,\Rightarrow\,\,\,\,\,\,\,\,\,\,\,\,x_{_{\mbox{\tiny max},out}}=0\,\,,
\end{equation}
and
\begin{equation}
\left[\frac{\partial  (\varphi_{_{out}}+ \psi_{_{out}}) }{\partial {k_{y}}}\right]_{_{(0,0)}} = 0\,\,\,\,\,\,\,\,\,\,\,\,\Rightarrow\,\,\,\,\,\,\,\,\,\,\,\,z_{_{\mbox{\tiny max}}}\,\cos 2\theta\, -\, y_{_{\mbox{\tiny max}}} \,\sin 2\theta= d\,\,,
\label{eq:eqd0}
\end{equation}
where
\begin{equation}
d= a\,(\cos\theta + \sin \theta)+ b\,\sin\theta\,\left(\frac{\cos\theta}{\sqrt{n^{\2} - \sin^{\2} \theta}}-1 \right)\,\,.
\label{eq:d}
\end{equation}
For refractive index $n=\sqrt{2}$, we can observe the output beam maximum moving along
\begin{equation}
\begin{array}{lcl}
 y = \displaystyle{\frac{b}{\sqrt{2}}\,\left( 1 - \frac{1}{\sqrt{2\,n^{\2}-1}}\right)}-a\,\sqrt{2}
 & & \mbox{for}\,\,\,\,\,\theta=\pi/4\,\,, \\
 z=a & & \mbox{for}\,\,\,\,\,\theta=0\,\,, \\
 y = \displaystyle{\frac{b}{\sqrt{2}}\,\left( 1 - \frac{1}{\sqrt{2\,n^{\2}-1}}\right)}
 & & \mbox{for}\,\,\,\,\,\theta=-\,\pi/4\,\,.
\end{array}
\end{equation}
For $\theta=\pi/4$, the choice of
\[ a= \frac{\sqrt{2\,n^{\2} - 1} - 1}{2\, \sqrt{2\,n^{\2} - 1}}\,\, b\,\,,\]
implies $y=0$. Consequently the incoming and outgoing beams move in the same direction.

\section*{\normalsize V. THE GOOS-H\"ANCHEN SHIFT BY THE SPM}

The optical path obtained in the previous section by using the SPM can be also determined by applying the Snell law. In this section, we present  the calculation of the additional quantum phase which cannot be predicted  by geometrical optics. At the down interface, the reflection coefficients for $s$ and $p$ polarized wave are
\[  \left\{\,r_{_{*}}^{^{(s)}}\,,\,r_{_{*}}^{^{(p)}}\,\right\}    = \left\{\,\frac{q_{z_{_{*}}}-\,k_{z_{_{*}}}}{q_{z_{_{*}}}+\,k_{z_{_{*}}}}\,,\,    \frac{q_{z_{_{*}}}-\,n^{\2}k_{z_{_{*}}}}{q_{z_{_{*}}}+\,n^{\2}k_{z_{_{*}}}}\,\right\}\,    \,\exp[\,2\,i\,q_{z_{_{*}}} a_{_{*}}]\,\,.
\]
Let us expand $k_{z_{_{*}}}$ around the center of the gaussian wave number distribution, i.e. $k_x=k_y=0$,
\begin{eqnarray}
k_{z_{_{*}}}^{^{2}}(k_{x}, k_{y}) & = &k_{z_{_{*}}}^{^{2}}(0,0)\, +\, \left[\,\frac{\partial k_{z_{_{*}}}^{^{2}}}{\partial k_{y}}\,\right]_{_{(0, 0)}}\,\hspace*{-0.5cm} k_y\, +\, O[k_{x}^{^{2}},k_{y}^{^{2}}]\nonumber \\
 & = &   k_{z_{_{*}}}^{^{2}}(0,0) \,+ \,2  \,q_{z_{_{*}}}(0,0) \left[\frac{\partial q_{z_{_{*}}}}{\partial k_{y}}\right]_{_{(0, 0)}} \hspace*{-0.5cm} k_y \, +\, O[k_{x}^{^{2}},k_{y}^{^{2}}]\,\,.
\label{eq:eqKzs02}
\end{eqnarray}
By using
\begin{equation}
k_{z_{_{*}}}^{^{2}} (0,0)= -\,\frac{\,\,k^{^{2}}}{2} \left(\,n^{\2}-2+2\sin\theta\sqrt{n^{\2}-\sin^{\2}\theta}\,\right) \nonumber
\,\,\,,\,\,\,\,\,\,\,
q_{z_{_{*}}}(0,0) = \frac{k}{\sqrt{2}}\,\sqrt{-\sin\theta+\sqrt{n^{\2}-\sin^{\2}\theta}}\,\,, \nonumber
\label{eq:kzs0dqzs}
\end{equation}
and
\begin{equation}
\left[\,\frac{\partial q_{z_{_{*}}}}{\partial k_{y}}\,\right]_{_{(0,0)}}=-\, \frac{\cos\theta}{\sqrt{2}} \left(1+\frac{\sin \theta}{ \sqrt{n^2 - \sin^2 \theta}} \right)\,\,, \nonumber
\end{equation}
we obtain
\begin{equation}
\frac{k_{z_{_{*}}}^{^{2}}(k_{x}, k_{y})}{k^{^{2}}} = \left( 1- \frac{\,\,n^{\2}}{2} - \sin\theta\sqrt{n^{\2}-\sin^{\2}\theta}\,\right)
  - \frac{\cos\theta \,(\,n^{\2} - 2\,\sin^{\2}\theta\,)}{\sqrt{n^{^{2}}-\sin^{\2}\theta}}\,\,\frac{k_{y}}{k}\,\,.
\label{eq:eqkzs2b}
\end{equation}
For $k_{y}\,>\,\sigma(n,\theta)\,k$,
\begin{equation}
\sigma(n,\theta) = \frac{\left(1-\displaystyle{\frac{\,\,n^{\2}}{2}}-\sin\theta \sqrt{n^{\2}-\sin^{\2}\theta}\right) \sqrt{n^{^2}-\sin^{\2}\theta}}{\cos\theta\left(n^{\2}-2 \sin^{\2}\theta\right)}\,\,,
\label{eq:sigma}
\end{equation}
we have $k_{z_{_*}}^{^{2}}<0$ and consequently an additional quantum phase has to be considered in calculating the optical path,
\begin{eqnarray}
\left\{\,\widetilde{\psi}^{^{\,(s)}}_{_{out}}\,,\,\widetilde{\psi}^{^{\,(p)}}_{_{out}}\,\right\}
&=&\left\{\,
\mbox{Arg}\left[\,\frac{q_{z_{_*}}-\,i\,|k_{z_{_*}}|}{q_{z_{_*}}+\,i\, |k_{z_{_*}}|} \,\right]\,,\, \mbox{Arg}\left[\,\frac{q_{z_{_*}}-\,i\,n^{\2}|k_{z_{_*}}|}{q_{z_{_*}}+\,i\, n^{\2}|k_{z_{_*}}|} \,\right]\right\}\nonumber \\
&=&-\,2\,\left\{\,\arctan\left[\,\frac{|k_{z_{_*}}|}{\,\,q_{z_{_*}}}\,\right]\,,\,
\arctan\left[\,\frac{n^{\2}|k_{z_{_*}}|}{\,\,q_{z_{_*}}}\,\right]
\,\right\}\,\,.
\end{eqnarray}
The derivatives of these phases,
\begin{equation}
\left\{\,\frac{\partial \widetilde{\psi}^{^{\,(s)}}_{_{out}}}{\partial {k_{y}}}\,,\,\frac{\partial \widetilde{\psi}^{^{\,(p)}}_{_{out}}}{\partial {k_{y}}}\,\right\} = \,\,\,\,\,\frac{2}{|k_{z_{_{*}}}|}\,\frac{\partial q_{z_{_{*}}}}{\partial {k_{y}}}\,\left\{\,1\,,\,\frac{n^{\2}\,k^{^{2}}}{k^{^{2}}\,+(n^{\2}+1)\,|k_{z_{_{*}}}|^{^{2}}}\,\right\}\,\,,
\end{equation}
will be then used to obtain the GH shift. To determine at which $k_y$-value the previous derivatives have to be calculated, we analyze the wave number distribution for different incidence angle, see Fig.\,2. For $\,\sigma(\sqrt{2},\theta)\,k\,\mbox{w}_{\0}\leq -5$ (Fig.\,2a), the wave packet is totally reflected and its wave number distribution is a symmetric distribution. Thus, the derivative of additional phase must be calculated at its center, $k_y=0$. For $\,\sigma(\sqrt{2},\theta)\,k\,\mbox{w}_{\0}\leq 5$ (Fig.\,2e), the reflection coefficient is real and there is not an additional phase. The intermediate case, $\,\sigma(\sqrt{2},\theta)\,k\,\mbox{w}_{\0}=0$ (Fig.\,2c), represents incidence at critical angle. In this case, only an half part of the wave number distribution contains an additional phase. Consequently, the derivatives have to be calculated at $k_y=k_c$,
\begin{equation}
k_c=\frac{\mbox{\Large{$\int$}}_{_{0}}^{^{+\,\infty}}\hspace*{-.2cm}\mbox{d}k_{y} \,\,k_{y} \, \exp\left[\,-\,(k_y \mbox{w}_{\0})^{^2}/\,2 \,\right]}{
\mbox{\Large{$\int$}}_{_{-\infty}}^{^{+\,\infty}}\hspace*{-.2cm}\mbox{d}k_{y} \,\, \exp\left[\,-\,(k_y \mbox{w}_{\0})^{^2}/\,2 \,\right]} = \frac{1}{\sqrt{2\, \pi}\,\, \mbox{w}_{\0}}\,\,.
\label{kc}
\end{equation}

\noindent $\bullet$ \fbox{$\sigma(n,\theta)\,k\,\mbox{w}_{\0}\leq -5$}\\

\noindent
In this case, we have to calculate the derivatives in $(k_x,k_y)=(0,0)$. We find
\begin{equation}
\left\{\,\frac{\partial \widetilde{\psi}^{^{\,(s)}}_{_{out}}}{\partial {k_{y}}}\,,\,\frac{\partial \widetilde{\psi}^{^{\,(p)}}_{_{out}}}{\partial {k_{y}}}\,\right\}_{_{(0,0)}} =\,\, -\, \left\{\, \widetilde{d}_{_{T}}^{^{\,\,(s)}}\,,\,\,\,\widetilde{d}_{_{T}}^{^{\,\,(p)}}\,\right\}\,\,,
\label{eq:eqPT}
\end{equation}
with
\begin{eqnarray}
\widetilde{d}_{_{T}}^{^{\,\,(s)}} & = &
 \frac{2\,\cos\theta}{k\,\sqrt{n^{\2}-2+2\sin\theta\sqrt{n^{\2}-\sin^{\2}\theta}}}\,\left(1+\frac{\sin \theta}{ \sqrt{n^2 - \sin^2 \theta}} \right)\,\,,\nonumber \\
\widetilde{d}_{_{T}}^{^{\,\,(p)}} & = &
\frac{n^{\2}}{1+ (n^{\2}+1)\left(\displaystyle{\frac{n^{\2}}{2}} - 1 + \sin\theta \sqrt{n^{\2}-\sin^{\2}\theta}\right)} \,\widetilde{d}_{_{T}}^{^{\,\,(s)}}\,\,.
\label{eq:dTpt}
\end{eqnarray}

\noindent $\bullet$ \fbox{$\sigma(n,\theta)\,k\,\mbox{w}_{\0}=0$}\\

\noindent
For incidence at critical angle, we have to calculate the derivatives in $(k_x,k_y)=(0,k_c)$, i.e.
\begin{equation}
\left\{\,\frac{\partial \widetilde{\psi}^{^{\,(s)}}_{_{out}}}{\partial {k_{y}}}\,,\,\frac{\partial \widetilde{\psi}^{^{\,(p)}}_{_{out}}}{\partial {k_{y}}}\,\right\}_{_{(0,k_c)}} = \,\,\,\,\,\frac{2}{|k_{z_{_{*}}}(0,k_c)|}\,\left[\frac{\partial q_{z_{_{*}}}}{\partial {k_{y}}}\right]_{_{(0,k_c)}}\left\{\,1\,,\,
\frac{n^{\2}\,k^{^{2}}}{k^{^{2}}\,+(n^{\2}+1)\,|k_{z_{_{*}}}(0,k_c)|^{^{2}}}\,\right\}\,\,.
\end{equation}
Recalling that at critical angles $k_{z_{_{*}}}(0,0)=0$, by using Eq.\,(\ref{eq:eqKzs02}),
\begin{equation}
k_{z_{_{*}}}^{^{2}}(0,\left\langle k_y\right\rangle) \,\,\approx\,\, 2\,k\,\sqrt{n^{\2}-1}\,\left[\,\frac{\partial q_{z_{_{*}}}}{\partial k_{y}}\,\right]_{_{(0,0)}}\,\hspace*{-0.5cm}\left\langle k_y\right\rangle
\end{equation}
and observing that
\begin{equation}
\left[\,\frac{\partial q_{z_{_{*}}}}{\partial {k_{y}}}\,\right]_{_{(0,k_c)}}\,\, \approx\,\, \left[\frac{\partial q_{z_{_{*}}}}{\partial {k_{y}}}\right]_{_{(0,0)}}\,\,,
\end{equation}
we get
\begin{equation}
\left\{\,\frac{\partial \widetilde{\psi}^{^{\,(s)}}_{_{out}}}{\partial {k_{y}}}\,,\,\frac{\partial \widetilde{\psi}^{^{\,(p)}}_{_{out}}}{\partial {k_{y}}}\,\right\}_{_{(0,k_c)}} =
\,\, -\, \left\{\, \widetilde{d}_{_{C}}^{^{\,\,(s)}}\,,\,\,\,\widetilde{d}_{_{C}}^{^{\,\,(p)}}\,\right\}\,\,,
\end{equation}
with
\begin{eqnarray}
\widetilde{d}_{_{C}}^{^{\,\,(s)}} & = &
\sqrt{2} \,\,\frac{\sqrt{k\,\mbox{w}_{\0}}}{k}\, \left[\,\frac{\pi}{\,(n^{\2}-1)}\,\right]^{^{1/4}}\,
\sqrt{ \cos\theta_{_C}\,\left(1+\frac{\sin \theta_{_C}}{ \sqrt{n^2 - \sin^2 \theta_{_C}}} \right)}\,\,,\nonumber \\
\widetilde{d}_{_{C}}^{^{\,\,(p)}} & = &   n^{\2} \,\widetilde{d}_{_{C}}^{^{\,\,(s)}}\,\,.
\label{eq:dCst}
\end{eqnarray}
The analytical formulas obtained for the GH shift  will be tested in the next section by a numerical analysis.

\section*{\normalsize VI. NUMERICAL ANALYSIS}

In this section, we present a numerical analysis of the GH shift for gaussian optical beams. The intensity of the outgoing beam is given by
\begin{eqnarray}
\mbox{I}_{_{out}}(\boldsymbol{r}) & = &  \left|\,E_{_{out}}(\boldsymbol{r})\,\right|^{^{2}} \\
& = & \mbox{I}_{_{0}}\,\,\left|\,\frac{\mbox{w}_{\0}^{\2}}{4 \pi}\,\int\hspace*{-.1cm}
\mbox{d}k_x\,\mbox{d}k_y\,\, \left|\,t^{^{(s,p)}}\right|\, \exp \left[-\frac{(k_{x}^{^2} +
k_{y}^{^2})\,\mbox{w}_{\0}^{\2}}{4}\,+\,i\,\left(\,\boldsymbol{k}_{_{out}}\cdot\,\, \boldsymbol{r}_{_{out}}\,+\,\,\psi_{_{out}}\,+\,\,\widetilde{\psi}^{^{\,(s,p)}}_{_{out}}\,\right)\,\right]\,
\right|^{^{2}}\,\,.\nonumber
\label{eq:eqteste}
\end{eqnarray}
To estimate the GH shift, we calculate the deviation from the geometrical maximum,
\[  z_{_{\mbox{\tiny max}}} \cos 2\theta - y_{_{\mbox{\tiny max}}} \sin 2\theta = d \,\,,\]
given in section IV (and also obtained from the Snell law).  In Fig.\,3, we plot the numerical data corresponding to the GH shift,
for $s$ and $p$ polarized waves, obtained for a fixed refractive index, $n=\sqrt{2}$, by varying the incidence angle and $k\,\mbox{w}_{_{0}}(=30,50,500)$. In Fig.\,4, the plots refer to a fixed incidence angle, $\theta=0$, and a varying refractive index. The numerical analysis shows an excellent agreement with our analytical prediction for the shift at critical angles. Observe that
\[  \sqrt{\frac{k}{\mbox{\,\,w}_{\0}}}\,\, \widetilde{d}_{_{C}}^{^{\,\,(s,p)}}\]
only depends on the refractive index $n$ (note that $\theta_c$ can be expressed as a function of $n$),
see Eqs.\,(\ref{eq:dCst}).

For $\sigma(n,\theta)\,k\,\mbox{w}_{\0}\leq -5$, we have to use for the GH shift the analytical expressions given in Eqs.\,(\ref{eq:dTpt}). Now,
\[  \sqrt{\frac{k}{\mbox{\,\,w}_{\0}}}\,\, \widetilde{d}_{_{T}}^{^{\,\,(s,p)}}\]
is proportional to $1/\sqrt{k\,\mbox{w}_{_{0}}}$ (in Figs.\,3 and 4, we have used $k\,\mbox{w}_{_{0}}=500$ for the analytical curve).
For a fixed refractive index, say $n=\sqrt{2}$,
\[
\sigma(\sqrt{2},\theta)\,k\,\mbox{w}_{\0}\,\,\leq\,\, -5\,\,\,\,\,\,\,\Rightarrow\,\,\,\,\,\,\,
\tan\theta \,\,\frac{2-\sin^{\2}\theta}{2\,\cos^{\2}\theta}\,\,\geq\,\, \frac{5}{k\,\mbox{w}_{\0}}\,\,.
\]
For such angles, we can use our analytical expression which shows an excellent agreement with the numerical data, see Fig.\,3.  For a fixed incidence angle, say $\theta=0$,
\[
\sigma(n,0)\,k\,\mbox{w}_{\0}\,\,\leq\,\, -5\,\,\,\,\,\,\,\Rightarrow\,\,\,\,\,\,\,
\frac{n^{\2} -2}{2\,n}\,\,\geq\,\, \frac{5}{k\,\mbox{w}_{\0}}\,\,.
\]
For such refractive indexes, the analytical expression  shows an excellent agreement with the numerical data, see Fig.\,4. We conclude this section, by observing that the numerical data confirms a maximum GH shift  for incidence angles {\em greater} than the critical ones\cite{1986JOSAA3}.

\section*{\normalsize VII. CONCLUSIONS}

In this paper, we have discussed  the frequency crossover for the GH shift. The analytical formulas obtained by the SPM have been then numerically tested. Our study  sheds new light on the validity region of the analytical formulas and on the transition region between partial and total reflection. The geometrical optical path derived by the SPM in section IV can be also obtained by  the Snell law in geometrical optics.
Nevertheless, the GH shift cannot be predicted by geometrical optics. For the calculation of an analytical formula for this shift, the use of the SPM is not a matter of taste. It is important to observe that, for total reflection, our wave number distributions are symmetric and consequently the derivatives have to be calculated at their center, located at $k_y=0$. This lead to a shift of the order of  $\lambda/2\pi = c/\omega$.

In our analysis, the SPM has been extended to critical regime. The formula for critical incidence has been obtained by calculating the derivatives at $k_y=k_c$, see Eq.\,(\ref{kc}). This is due to the fact that for incidence at critical angles the wave number distribution is not symmetric. The GH shift is now amplified by the factor $\sqrt{k\,\mbox{w}_{\0}}$.

The numerical data allow to analyze the frequency crossover for the GH shift. They show an excellent agreement with our analytical predictions.  The data also show that the maximum GH shift is near to the critical angle\cite{1986JOSAA3}. Thus, experiments in this region represent the most favorable situation to investigate this shift.

In a forthcoming paper, we aim to analyze in detail the behavior of gaussian optical beams incident at critical angles in the case in which the outgoing wave number distribution is asymmetric.

\vspace{.8cm}

\noindent
\textbf{\footnotesize ACKNOWLEDGEMENTS}\\
 We gratefully thank  the Capes (M.\,P.\,A.), Fapesp (S.\,A.\,C.), and CNPq (S.\,D.\,L.)   for  the financial support and the referee for his useful suggestions and for drawing our attention to the references \cite{REF1,REF2}.

\newpage

\begin{figure}
%\centering
\vspace*{-3cm}
\hspace*{-1.7cm}
		\includegraphics[width=17cm, height=25.5cm, angle=0]{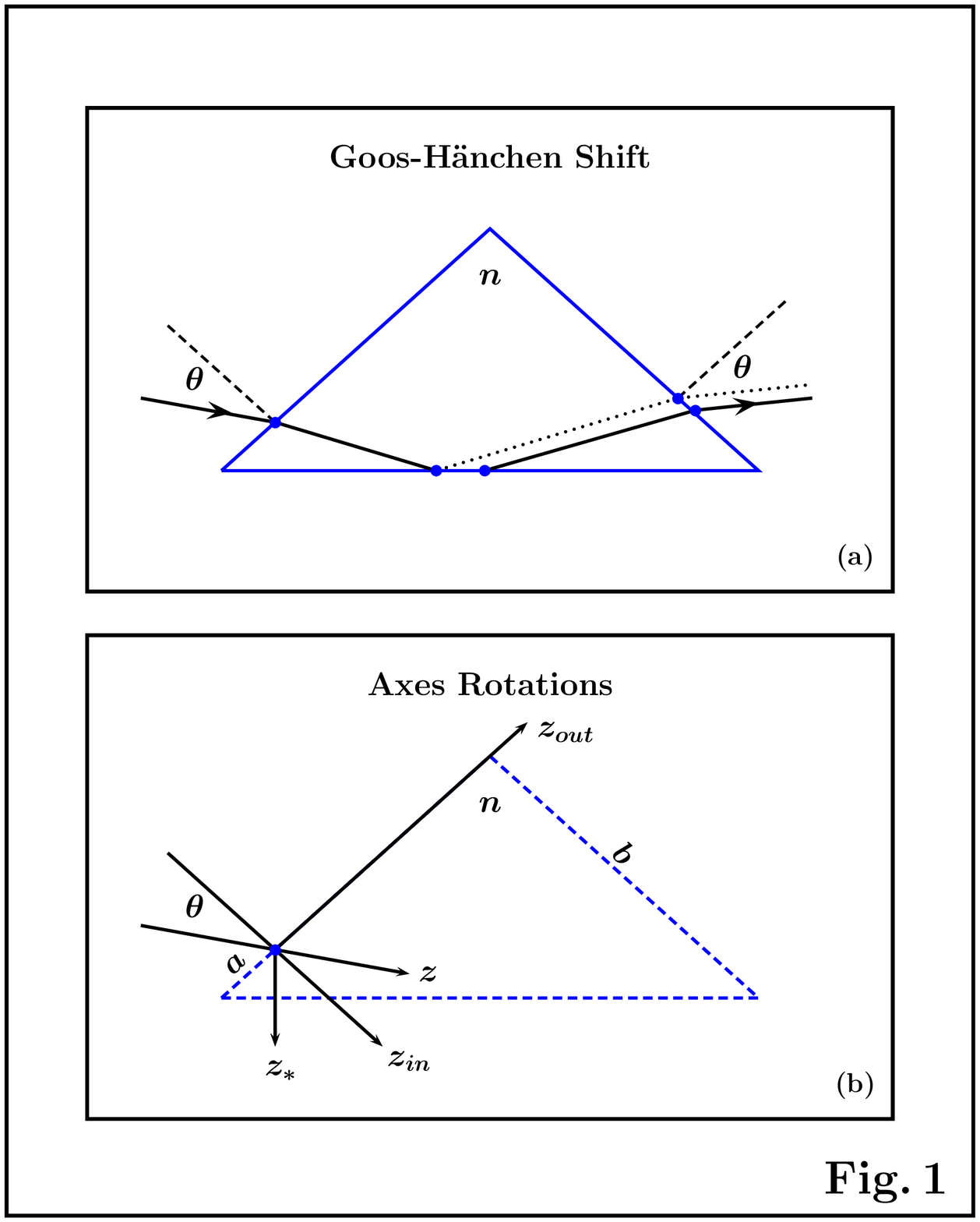}
\vspace*{-4cm}
\caption{Schematic diagram of the dielectric block analyzed in this paper. In (a), it is shown
the lateral displacement (solid line) of the reflected beam at the down dielectric-air interface with respect  to geometrical path (dotted line).  In (b), we draw the axes of the incoming propagation, $z$, and
of the normal to the left air/dielectric boundary, $z_{_{in}}$, to the down dielectric/air boundary, $z_{_*}$, and to the right air/dielectric boundary, $z_{_{out}}$.}
\label{fig:fig1}
\end{figure}

\newpage

\begin{figure}
%	\centering
		\vspace*{-3cm}
        \hspace*{-1.7cm}
		\includegraphics[width=17cm, height=25.5cm, angle=0]{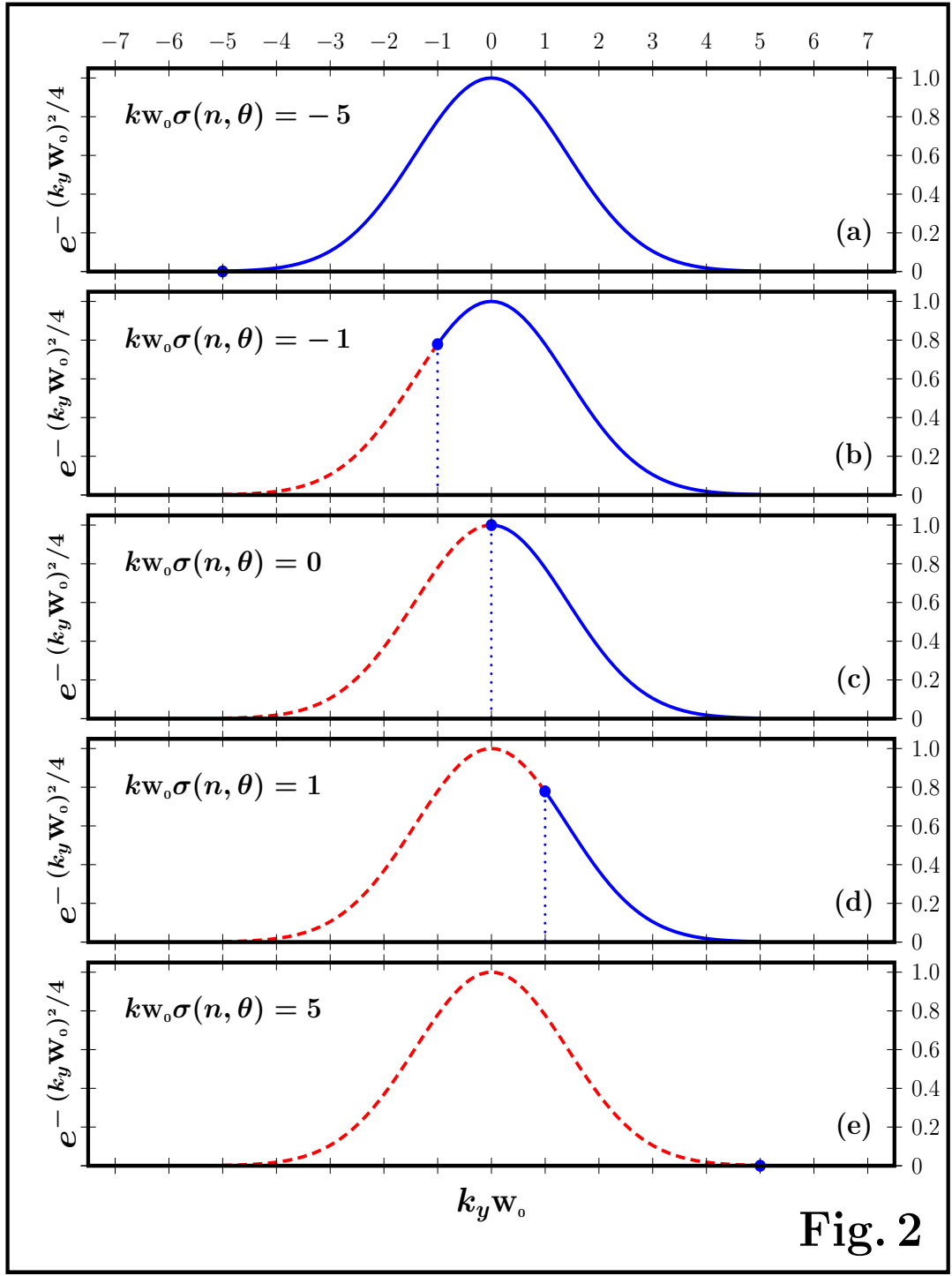}
		\vspace{-4cm}
		\caption{Gaussian wave number distribution for different incidence angles. The dotted line represents the part of the distribution with a real reflection coefficient, $k_{z_*}^{\2}>0$. The solid line
 represents the part of the distribution with a complex reflection coefficient, $k_{z_*}^{\2}<0$.}
	\label{fig:fig2}
\end{figure}

\newpage

\begin{figure}
%	\centering
\vspace*{-3cm}
\hspace*{-1.5cm}
		\includegraphics[width=17cm, height=25.5cm, angle=0]{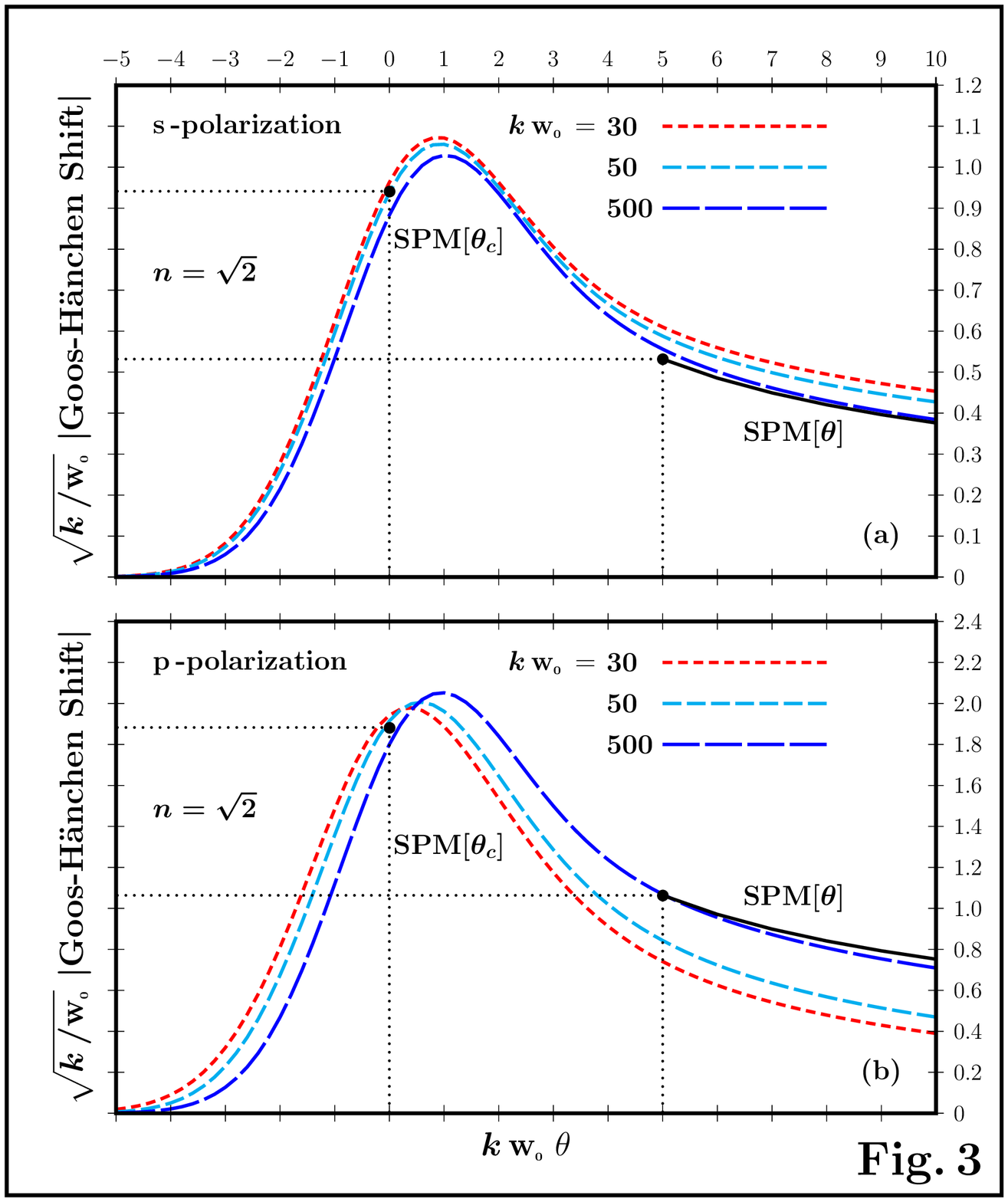}
		\vspace{-4cm}
		\caption{The numerical GH shift  is plotted as a function of the incidence angle for  a fixed refractive index,  $n=\sqrt{2}$, and  three different  values of $k \mbox{w}_{\0}$ (dashed lines). The numerical data are in excellent agreement with our analytical predictions for the GH shift, $\widetilde{d}_{_C}^{^{(s,p)}}$ (dot) and $\widetilde{d}_{_T}^{^{(s,p)}}$ (solid line for
$k \mbox{w}_{\0}=500$).}
	\label{fig:fig3}
\end{figure}

\newpage

\begin{figure}
	%\centering
 \vspace*{-3cm}
\hspace*{-1.5cm}
		\includegraphics[width=17cm, height=25.5cm, angle=0]{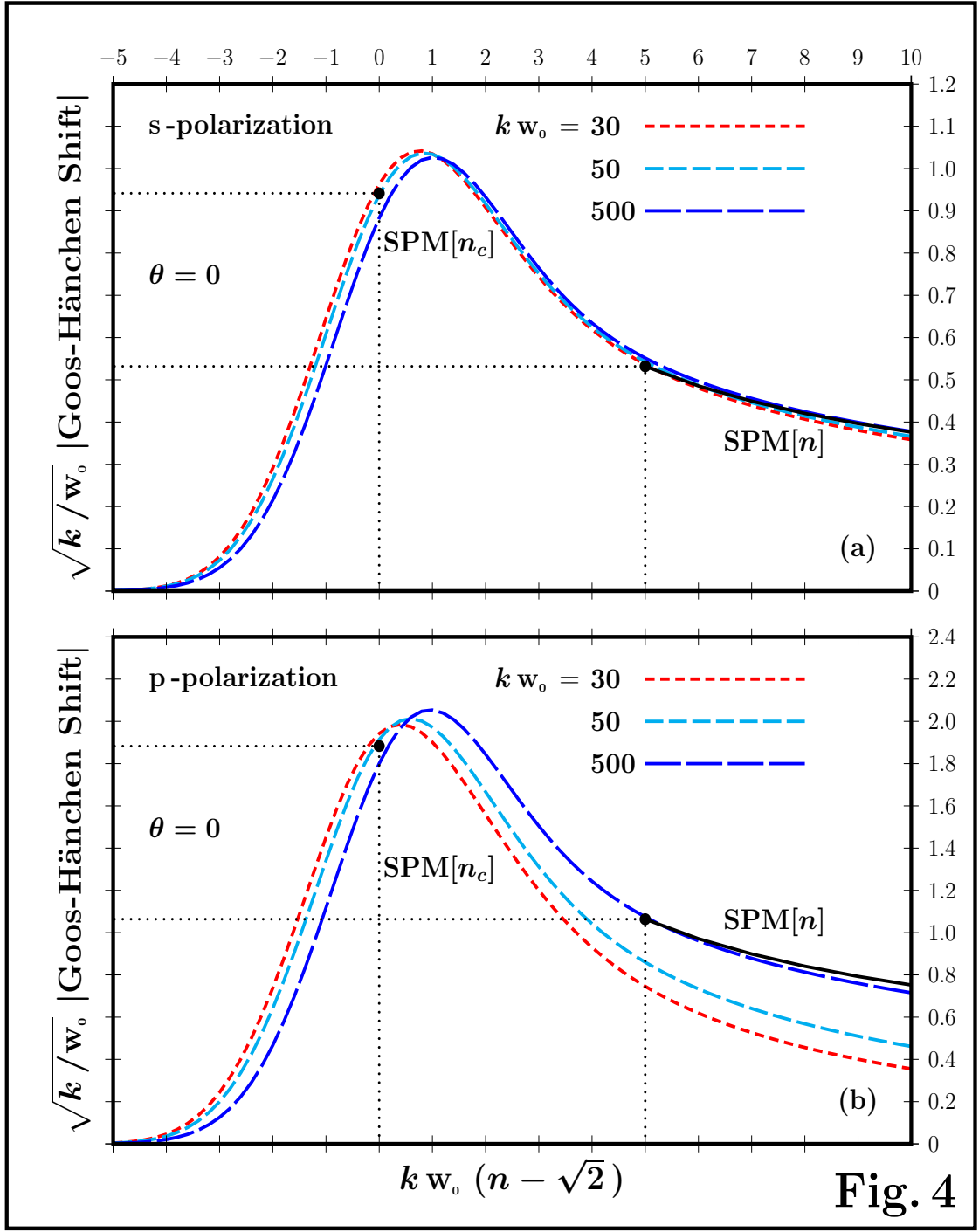}
		\vspace{-4cm}
		\caption{The numerical GH shift  is plotted as a function of the refractive index  for  a fixed incidence angle,  $\theta=0$, and  three different  values of $k \mbox{w}_{\0}$ (dashed lines). The numerical data are in excellent agreement with our analytical predictions for the GH shift, $\widetilde{d}_{_C}^{^{(s,p)}}$ (dot) and $\widetilde{d}_{_T}^{^{(s,p)}}$ (solid line for
$k \mbox{w}_{\0}=500$).}
	\label{fig:fig4}
\end{figure}

\end{document}